\documentclass[reprint, superscriptaddress, amsmath,amssymb, aps, pra]{revtex4-1}

\usepackage{color}

\usepackage{siunitx}
\usepackage{natbib}
\usepackage{amsmath}
\usepackage{tikz}
\usepackage{pifont}
\usepackage{wasysym}

\usepackage{graphicx}
\usepackage{dcolumn}
\usepackage{bm}
\usepackage{siunitx}
\usepackage{upgreek}

\DeclareSIUnit{\rad}{rad}

\begin{document}

\preprint{APS/123-QED}

\title{Trapping microparticles in a structured dark focus}

\author{F. Almeida}
\email{felipealmeida@aluno.puc-rio.br}
\affiliation{Departamento de F\'{i}sica, Pontif\'{i}cia Universidade Cat\'{o}lica do Rio de Janeiro,  22451-900 Rio de Janeiro, RJ, Brazil}
\author{I. Sousa}
\affiliation{Departamento de F\'{i}sica, Pontif\'{i}cia Universidade Cat\'{o}lica do Rio de Janeiro,  22451-900 Rio de Janeiro, RJ, Brazil}
\author{O. Kremer}
\affiliation{Centro de Estudos em Telecomunicações, Pontif\'{i}cia Universidade Cat\'{o}lica do Rio de Janeiro,  22451-900 Rio de Janeiro, RJ, Brazil}
\author{B. Pinheiro da Silva}
\affiliation{Instituto de Física, Universidade Federal Fluminense, Niterói, Rio de Janeiro 24210-346, Brazil}
\author{D.~S.~Tasca}
\affiliation{Instituto de Física, Universidade Federal Fluminense, Niterói, Rio de Janeiro 24210-346, Brazil}
\author{A. Z. Khoury}
\affiliation{Instituto de Física, Universidade Federal Fluminense, Niterói, Rio de Janeiro 24210-346, Brazil}
\author{G. Tempor\~ao}
\affiliation{Centro de Estudos em Telecomunicações, Pontif\'{i}cia Universidade Cat\'{o}lica do Rio de Janeiro,  22451-900 Rio de Janeiro, RJ, Brazil}
\author{T. Guerreiro}
\email{barbosa@puc-rio.br}
\affiliation{Departamento de F\'{i}sica, Pontif\'{i}cia Universidade Cat\'{o}lica do Rio de Janeiro,  22451-900 Rio de Janeiro, RJ, Brazil}

\date{\today}


\begin{abstract}
We experimentally demonstrate stable trapping and controlled manipulation of silica microspheres in a structured optical beam consisting of a dark focus surrounded by light in all directions - the Dark Focus Tweezer. Results from power spectrum and potential analysis demonstrate the non-harmonicity of the trapping potential landspace, which is reconstructed from experimental data in agreement to Lorentz-Mie numerical simulations. Applications of the dark tweezer in levitated optomechanics and biophysics are discussed.
\end{abstract}

\maketitle

\textit{Introduction.} -- Light exerts forces upon matter \cite{Beth1936}. As shown by Arthur Ashkin \cite{ashkin1997}, these forces can be used to create stable traps for nano- and microscopic dielectric particles, with a myriad of applications from fundamental physics \cite{delic2020cooling, windey2019cavity, magrini2021real, rieser2022observation, Blakemore2022, Afek2022} to metrology \cite{Barzanjeh2022, ricci2022chemical, asano2022}, quantum information \cite{Fiaschi2021,Houhou2022} and biology \cite{nussenzveig2018, araujo2019, bustamante2021}. When the refractive index of the particle's material is larger than that of its surrounding medium, optical forces attract the object towards high intensities of light. For Gaussian beam optical tweezers, the resulting potential is approximately harmonic \cite{Suassuna2021}, and careful calibration of the trap by a number of different methods \cite{perez2018high, gieseler2021optical} allows for precision force microscopy down to the molecular realm \cite{bustamante2022development}. 

A growing interest in the fields of levitated optomechanics and optical micro-manipulation is in enhanced force effects due to structured materials and light beams. For example, stable Casimir trapping of refractive-index engineered materials \cite{zhao2019stable}, enhanced forces in optically active nano-crystals \cite{shan2021optical} and nitrogen-vacancy colour center ensembles \cite{juan2017cooperatively} and probing of structured beams using levitated nanorods \cite{hu2022structured} have been demonstrated, while enhanced optical tweezing of meta-atoms exploiting Mie-resonances \cite{lepeshov2023levitated}, composite microspheres \cite{Ali2020} and chiral sorting of microparticles proposed \cite{ali2020probing, ali2021enantioselection}. Within this context of engineered nano and micro-traps, we can also explore repulsive optical forces: in the situation that a particle has a lower refractive index than its surrounding medium, it gets expelled from high intensity regions of light \cite{ashkin1970acceleration}. Using structured beams \cite{yang2021optical}, we can then engineer an inverted optical trap -- a dark focal region surrounded by a bright barrier \cite{arlt2000} -- capable of trapping 
an object and shielding it from external influence. 


A Dark Focus Tweezer (DFT) could find many applications across physics and biology. The optical potential generated by structured light dark traps can have tunable non-harmonicity \cite{melo2020b}, providing a laboratory for studies of non-linear stochastic dynamics \cite{Suassuna2021} and non-Gaussian state preparation in optomechanics \cite{neumeier2022fast}. Moreover, trapping objects in the dark can be extremely beneficial in the fields of active matter and biophysics, where laser damage limits experiments with living cells \cite{zhang2008optical, ehrlicher2002guiding, BlzquezCastro2019}. 

In this letter we report the construction of a dark optical trap for microparticles as proposed and theoretically analysed in \cite{melo2020b}. Earlier experiments have employed structured light and optical bottle beams to manipulate atoms in blue-detuned lasers \cite{isenhower2009atom, xu2010trapping, barredo2020three, Zhang2011} and micron-sized objects through photophoretic and thermal forces \cite{gong2016controllable}. Here we demonstrate stable trapping and controlled manipulation of microparticles in a structured light dark focus through optical forces alone. As we will show, the DFT induces a strongly non-harmonic potential landscape reflected as non-Gaussianity in the statistical properties of the particle's stochastic trajectory. We probe the particle motion both through the its power spectrum density as well as potential analysis and reconstruct the optical potential landscape through matching of data with numerical simulations. 

\begin{figure*}[ht!]
    \centering
    \includegraphics[width=0.9\textwidth]{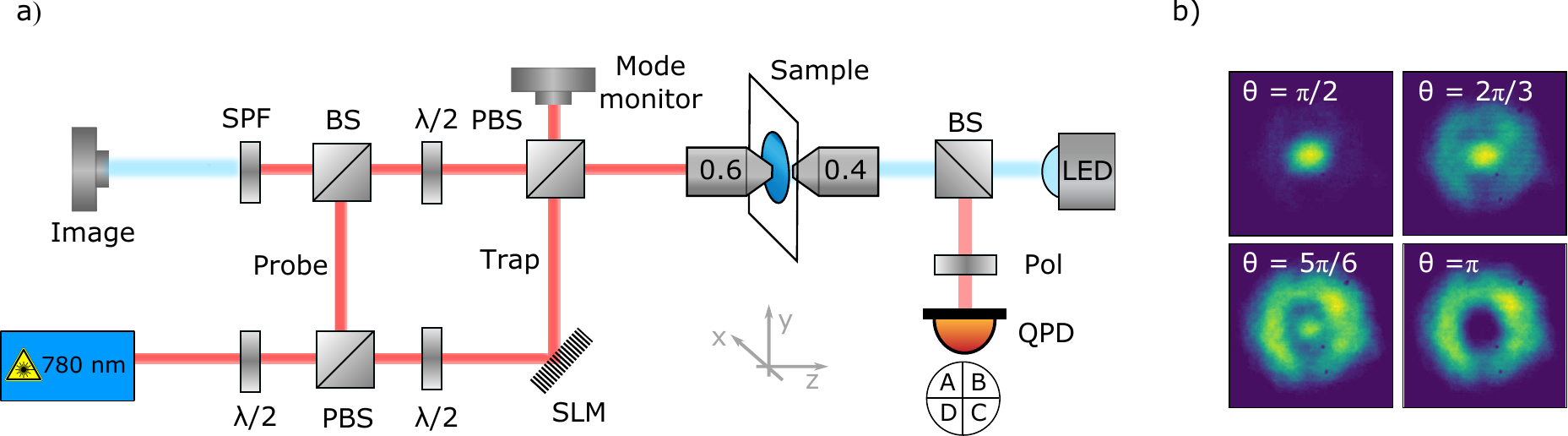}
    \caption{a) Simplified setup: a \SI{780}{nm} CW laser is split into orthogonal polarizations by a half wave plate ($\lambda/2$) and a polarizing beam splitter (PBS). The vertically polarized beam (Trap) is modulated by an SLM, and directed to an objective (NA $=0.6$) to generate the optical trap. The horizontal component (Probe) traverses the trap and is collected by a second objective (NA $=0.4$) and used to probe the motion of a trapped microparticle with a QPD. The trapping beam is filtered by a polarizer (Pol). Image of the trapped particle is obtained by focusing light from a LED onto the particle, subsequently collected by the trapping objective, filtered by a short pass filter (SPF) and focused onto a CCD (Image).
    b) Behavior of the intensity distribution around the focal point can be mimicked by varying the relative phase $ \theta $ between the Gaussian and LG modes.  
    } 
    \label{setup}
\end{figure*}

\textit{The dark focus tweezer.} -- The DFT, sometimes also called the optical bottle beam \cite{arlt2000}, consists of a dark focus surrounded by a bright intensity region \cite{melo2020b}. There are different ways of generating a DFT \cite{Yelin2004, Du2014, Wei2005, Yang2019, Whyte2005}; for simplicity, we choose a superposition of a Gaussian ($\ell = 0,  p = 0 $) with a Laguerre-Gauss beam ($\ell = 0,  p = 1 $) with a relative phase of $ \pi $ \cite{melo2020b}. This choice for a bottle beam allows for an intuitive description of the optical potential and can be readily generated using a spatial light modulator (SLM). 

For a DFT of wavelength $ \lambda_{0} $ in a medium of refractive index $ n_{m} $, the most important parameter is the numerical aperture $ \mathrm{NA}$, from which the beam waist $ \omega_0= \lambda_0 / \pi \mathrm{NA} $ and Rayleigh range $  z_R = n_{m}\lambda_0 / \pi \mathrm{NA}^2  $ can be calculated. The total intensity of the beam is $I_0=2P_0/\pi\omega_0^2$, where $ P_{0}$ is the total beam power.
Throughout this work, we adopt the intensity profile of a DFT with $ p = 1$. We can also define the width $ W $ and height $ H $ of the bottle as the distances between the peak values of intensity along the $ x $ and $ z $ directions; these are $W=2\omega_0,H=2z_R$, from which we see that the width scales as $ \mathrm{NA}^{-1} $ and height as $ \mathrm{NA}^{-2}$ (see supplemental material SM). 
 


For a particle of radius much smaller than the beam wavelength, $ R \ll \lambda_{0} $, optical forces due to a linearly polarized light beam are decomposed into scattering (non-conservative) and gradient (conservative) components, both increasing with the factor $ \alpha = \left[(m^2-1) / (m^2+2) \right] $, where $m=n_p/n_m$ is the particle-medium refractive index ratio. 
We are interested in situations where $ m < 1 $ (i.e. $ n_{p} < n_{m} $) and the particles are repelled by higher intensities of light \cite{grynberg2010}.
The gradient force field can be expressed in terms of the potential landscape,
\begin{eqnarray}
    \label{eq:potential_dipole}
 V(\vec{r})& =& -\frac{2\pi n_{m}R^3}{c}\alpha \ I(\vec{r})
 \label{eq:polynomial_potential}
\end{eqnarray}
where $I(\vec{r}) $ is the beam intensity at position $\vec{r}$.
Note that the potential may switch from attractive to repulsive depending on the value of $ m $; although our experimental conditions do not fit the dipole regime, this feature remains valid in our experiment.
As discussed in \cite{melo2020b}, near the origin (i.e. $\rho\ll\omega_0$, $z\ll z_R $), the potential $V(\vec{r})$ can be expanded as a polynomial function of coordinates,
\begin{eqnarray}\label{eq:potential-polynomial}
    V(\rho,z)\approx \frac{k_{z}}{2} z^2 - k_{\rho z} \rho^2z^2 + \frac{k_{\rho}}{4} \rho^4
    %
    ,
    \label{eq:potential_approx}
\end{eqnarray}
where $ k_{z} $ is the harmonic term strength along the axial direction and $ k_{\rho z},  k_{\rho} $ denote the anharmonic potential strengths.
In the Rayleigh regime, these potential coefficients are simple functions of the beam parameters (see SM)  \cite{melo2020b}.


In our experiment trapped particles have a radius of $ R = \SI{575}{nm}$, comparable to the wavelength $ \lambda_0 = \SI{780}{nm}$, a regime in which generalized Lorentz-Mie scattering theory must be employed for the calculation of optical forces \cite{jones2015}. 
Numerical simulation of the resulting force fields can be performed using the toolbox presented in \cite{nieminen2007}.
Eq. \eqref{eq:potential_approx} provides a good approximation to the potential landscape near the origin also in the intermediate regime, with root-mean-square deviations with respect to full numerical simulation of Lorentz-Mie scattering theory below 1\% for a wide range of particle radii (see SM).



\textit{Experimental setup.} -- 
The experimental setup for generating a dark focus tweezer can be seen in Fig. \ref{setup}a). A CW laser at \SI{780}{nm} (Toptica DL-pro) seeds a tapered amplifier (Toptica BoosTa) yielding \SI{1.6}{W} of power. The beam is divided by a half-wave plate and a polarizing beam splitter (PBS) to produce the trapping beam and an auxiliary probe beam.
The trapping beam is modulated by an SLM (Holoeye) and sent through an objective (Olympus UPlanFLN 100x adjustable NA $ = 0.6 - 1.3 $). The resulting superposition can be monitored in a camera providing a visualization of the beam's transverse profile as shown in Fig. \ref{setup}b). We change the relative phase between the modes in order to mimic the DFT's intensity pattern along the axial direction.
Image of the trapped particle is produced by focusing light from a LED into the sample, subsequently collecting it with the trapping objective and projecting onto a CCD (Image camera).

SiO$_{2}$ beads of radius $ R = \SI{575}{nm}$ (microParticles GmbH) with refractive index $ n_{p} = 1.45$ are immersed in clover oil solution, with a refractive index of $ n_{m} = 1.53 $ and measured transmission for \SI{780}{nm} of $ \eta_{\rm clover} = 85\% $. 
To load the trap, we position the center of the beam at the location of a nanoparticle and
abruptly turn on the dark focus tweezer.
The hydrophobic nature of the oil increases the tendency of the silica microspheres to aggregate, occasionally forming microdumbells in addition to single particles. 


\begin{figure}[ht!]
    \centering
    \includegraphics[width=0.45\textwidth]{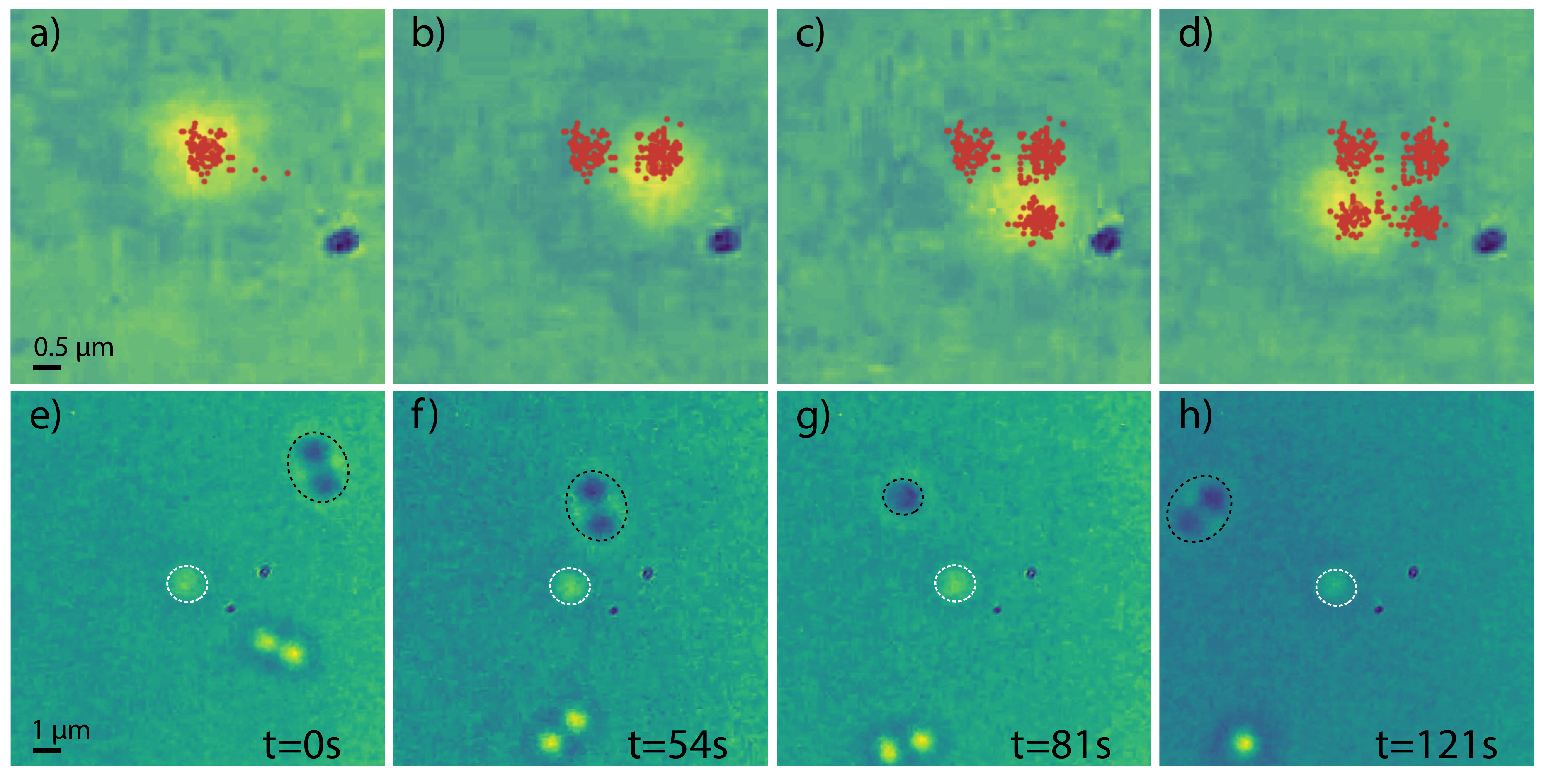}
    \caption{a)-d) Controlled SLM motion of the trapped particle. 
    Position is tracked by monitoring the coordinate of the brightest pixel. e)-h) Shielding effect: a dumbbell (black contour) is repelled after approaching a trapped particle (white contour). See supplementary videos.
    }
    \label{fig:OBB_effect}
\end{figure}

\textit{Controlled particle motion and shielding.} -- 
As a first demonstration of stable trapping in the DFT, we slightly move the beam by adjusting the SLM modulation angle allowing for a fine control of the particle position by deflecting the dark focus center. Fig. \ref{fig:OBB_effect}a-d) shows the iteration of four different trap positions, with red dots marking the brightest pixel in the image approximately corresponding to the center of the microsphere; see the supplementary video (S1).
These images are obtained by collecting the light from an LED scattered by the particle and registered with the CCD camera, as shown in Figure \ref{setup}.
By switching off the Gaussian component of the DFT superposition and producing a pure LG mode we observed the particle is lost from the trap.

In Gaussian optical tweezers, additional objects in the sample travelling nearby the trapped particle are drawn into the potential landscape by the attractive optical forces. In contrast, a particle trapped in the DFT is shielded from the influence of these external objects due to the repulsive optical force.
Fig. \ref{fig:OBB_effect}e)-h) displays typical subsequent frames of a trapped particle (white dotted circle) surrounded by free, passing-by particles. We observe a microdumbbell (black dashed circle) approaching the trapped particle and subsequently repelled by the DFT beam; see supplementary video (S2). 

\textit{Power spectrum analysis.} -- Among the most employed techniques to calibrate optical traps \cite{Gieseler2021, melo2020a} is the power spectrum density analysis (PSD). 
Analysis of the Langevin equation for a trapped particle in a harmonic potential reveals that the PSD has a Lorentzian form with the corner frequency parameter proportional to the trap's stiffness \cite{BergSrensen2004}. 

The potential associated to the DFT can be modeled by a fourth-order polynomial in the particle's coordinates, thus being non-harmonic (see SM). Numerical simulations of a trapped particle in the overdamped regime subject to quartic potentials show that the PSD of the particle motion is well fitted by a Lorentzian function, despite the exact relation between the corner frequency and the trap's strength parameters being unknown beyond perturbation theory \cite{Suassuna2021}. In effect, the PSD method cannot be directly used to determine the DFT's strength constants, but we can use it as a consistency check between numerical simulations of the particle motion subject to optical forces in the intermediate regime and experimental data. This indirectly provides information on the trap's characteristics. 

\begin{figure}[t!]
    \centering
    \includegraphics[width=0.4\textwidth]{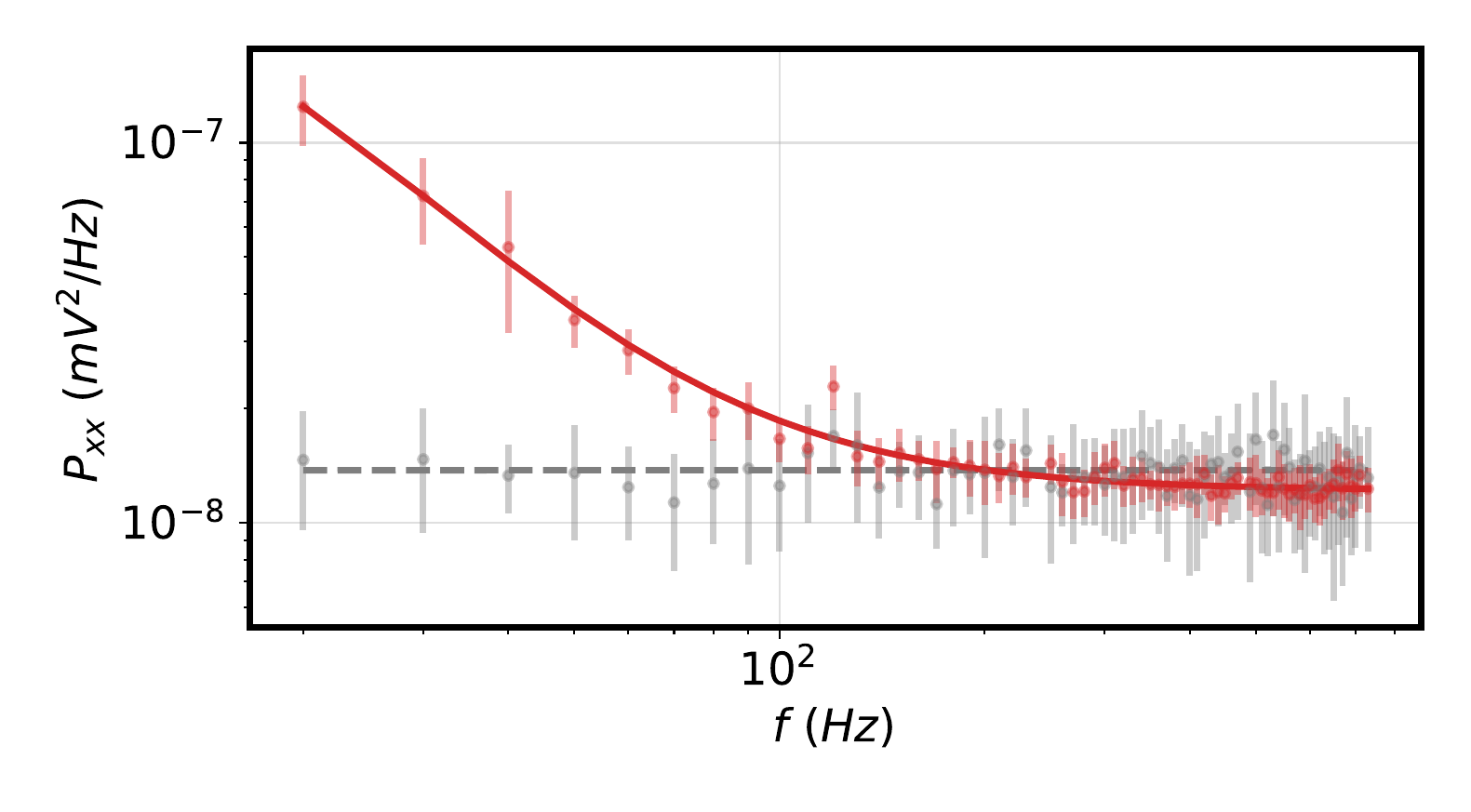}
    \caption{PSD of a particle trapped in the dark focus (red) in comparison to background noise (grey). The effective corner frequency is $f_{\mathrm {c, DFT}}=$ \SI[parse-numbers = false]{(13.4 \pm 0.7)}{Hz}. 
    }
    \label{fig:psd_section}
\end{figure}

Due to the nature of the dark trap, scattering of photons is greatly reduced, hindering motion detection by the traditional technique of collecting light scattered from the trapping beam \cite{tebbenjohanns2019}.
To overcome this we employ an auxiliary weak \textit{probe beam} in a Gaussian mode with polarization orthogonal to the trapping beam.
Being distinguishable and provided it has low power, the probe beam does not significantly alter the properties of the dark trap. 
Moreover, any eventual residual scattering noise due to the trapping beam can be filtered by a suitably aligned polarizer before detection, allowing access to the information carried by the probe alone.
The probe light scattered by the particle is collected by a second objective lens (Olympus PlanN 10x, NA = 0.25) and directed to a quadrant photodetector (QPD,  New Focus 2931) generating signals proportional to the particle's radial and axial coordinates (see SM for details). 

\begin{figure*}[t]
    \centering
    \includegraphics[width=\textwidth]{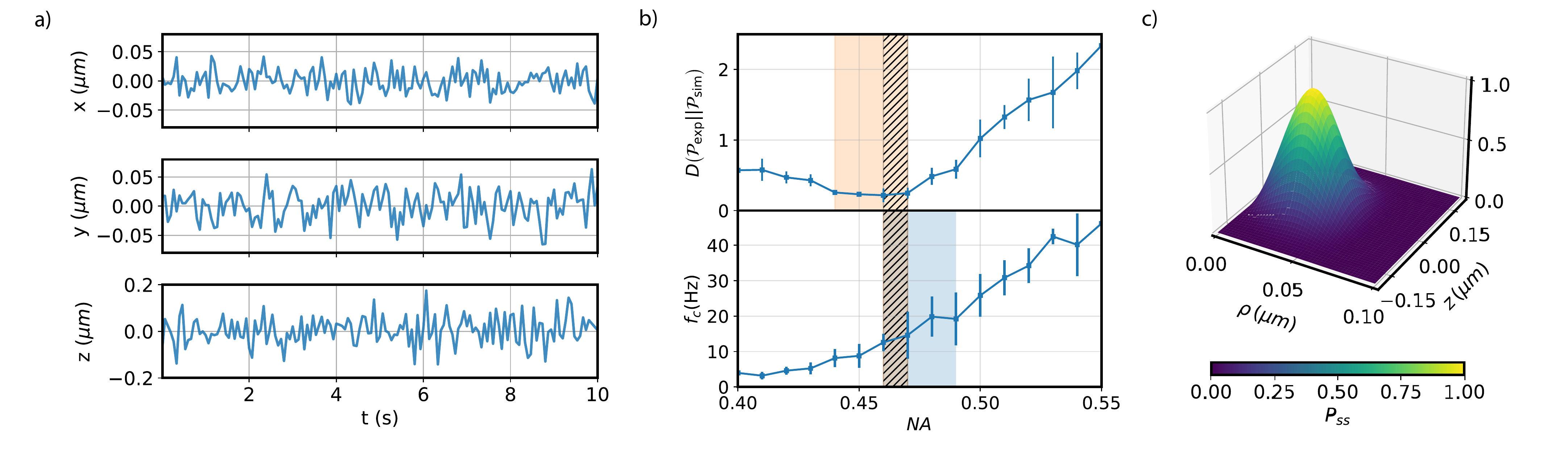}
    \caption{Potential analysis. a) Position traces of the particle. b) (Top) Kullback-Leibler divergence $ D(\mathcal{P}_{\rm exp} || \mathcal{P}_{\rm sim}) $ between simulation and experiment as functions of the NA. The orange region displays the range in which simulated and experimental probability distributions are most similar. (Bottom) Corner frequency of simulated dynamics. The blue region displays the range where the effective corner frequency of the simulation is within the error margin of measured value $f_{\mathrm {c, DFT}}=$ $\SI[parse-numbers = false]{(13.4 \pm 0.7)}{Hz}$. The hatched area shows the intersection of both methods, NA$=0.46 - 0.47$. c) Fitted normalized PDF of the centroid's position.
 }
    \label{fig:potential_analysis_painel}
\end{figure*}

We now turn to measurements of the PSD of a particle trapped in the dark focus. The measured PSD can be seen in Fig. \ref{fig:psd_section} (red dots and line) together with the background scattering noise in the absence of a trapped particle (grey dots and line), for comparison. A Lorentzian fit to the PSD yields an effective corner frequency of   $f_{\mathrm {c, DFT}}=$ \SI[parse-numbers = false]{(13.4 \pm 0.7)}{Hz}.
Numerical simulations of the trapped particle within the DFT suggests that this value lies in the NA range between $0.46 - 0.49$, where we find corner frequencies in the range between $f_{\mathrm {c, sim}}=$ \SI[parse-numbers = false]{(12.6 \pm 2.3)}{Hz} and $f_{\mathrm {c, sim}}=$ \SI[parse-numbers = false]{(19.2 \pm 7.3)}{Hz}, respectively.
Using the scaling of width $ W $ and height $ H $ with the NA introduced earlier, we estimate the order-of-magnitude of the trap size to lie between $\SI{1.0}{\mu m}$ and $\SI{1.1}{\mu m}$. Considering that our particles have a diameter of $\SI{1.15}{\mu m}$, this suggests the view that the trapped particle feels only loose forces within the dark focal region.

\begin{figure}[h]
    \centering
    \includegraphics[width=0.3\textwidth]{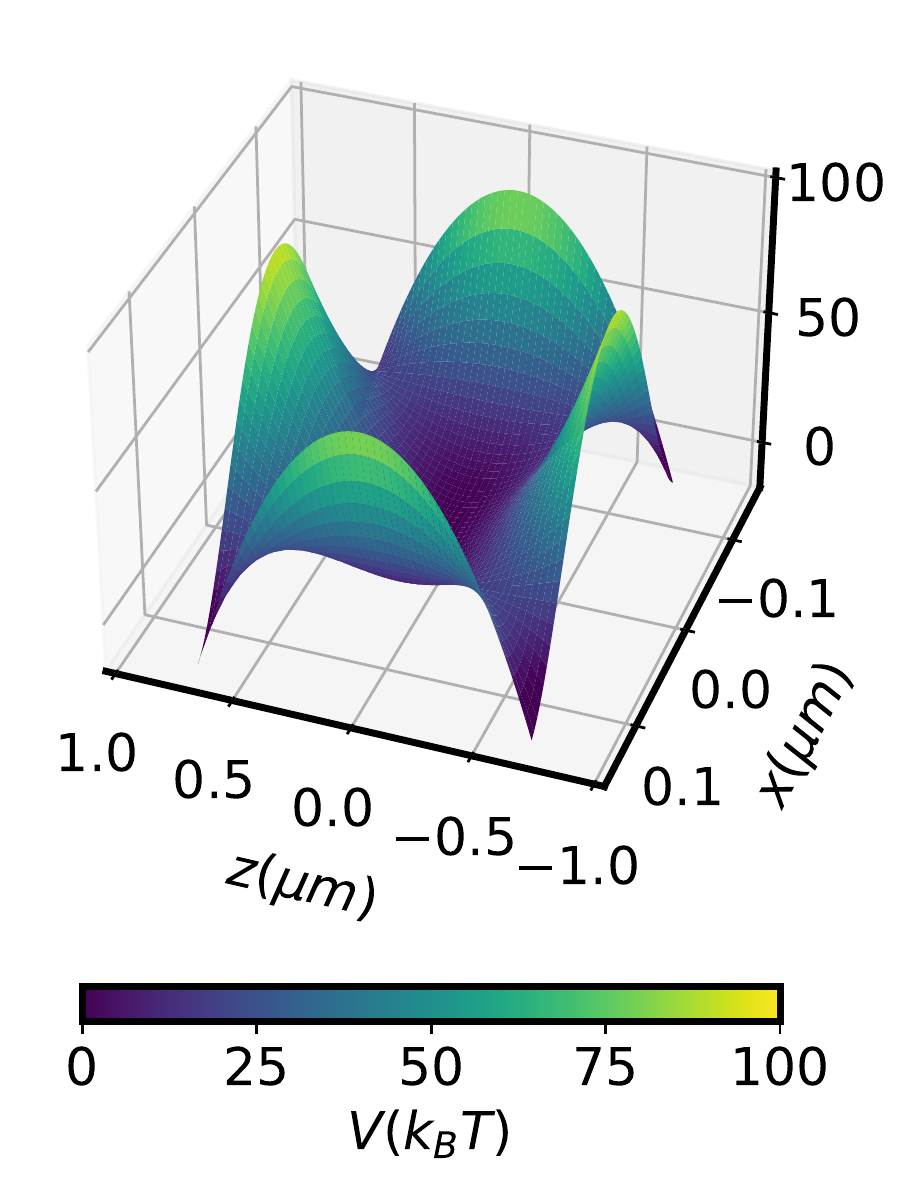}
    \caption{Reconstructed potential for centroid coordinate of a particle in the DFT. 
    }
    \label{fig:potential_recon}
\end{figure}




\textit{Potential analysis.} -- 
In thermal equilibrium and in the limit that the conservative force dominates over dissipative forces, the position probability density function (PDF) follows $ P(\rho,z) \propto \exp\left(  -V(\rho,z) / k_{B}T \right) $ 
and can be reconstructed from frames of the particle motion acquired with a CCD over long times \cite{Gieseler2021}.
We acquired long duration videos of a trapped particle at a rate of $15.0 \  \textrm{frames}/ \SI{}{s}$, from which we extract the particle's centroid and axial coordinates using image processing \cite{forsyth2002computer}. The resulting coordinate traces can be seen in Fig. \ref{fig:potential_analysis_painel}a).  Due to the potential anharmonicity, the position PDF is expected to be non-Gaussian; a Kolmogorov-Smirnov hypothesis test on the position time series confirms that at $0.05$ significance.

To find the best match between the data and the quartic potential model, we ran several simulations of the particle dynamics in the DFT parametrized by the trap's NA. We then extract simulated PDFs for motion along the transverse directions and numerically compute the Kullback-Leibler (KL) divergence between each of the simulated distributions and the marginal PDFs obtained from the experiment.
Minimizing the KL divergence between simulation and experiment is equivalent to performing a maximum likelihood estimation of the trap's NA \cite{rice2003mathematical}.
Fig. \ref{fig:potential_analysis_painel}b) displays the KL divergence averaged over the $ x $ and $y $ directions (top plot), where we find that a potential with NA = $ 0.46$ best describes the measured position traces, consistent with the expectation from Gaussian optics provided the beam waist prior to the SLM, given by NA $ \leq 0.6$.
Moreover, we compute the PSD of the numerical simulations' position data, from which we obtain the corner frequency of the Lorentzian fit. The obtained values of corner frequency for each simulation are plotted as a function of NA in Fig. \ref{fig:potential_analysis_painel}b) (bottom plot), where we see that NA = $0.46$ also displays the best agreement with the PSD measurements, $ f_{\mathrm {c, sim}} = \SI[parse-numbers = false]{(12.6 \pm 2.3)}{Hz} $. 

From the position traces we can reconstruct the PDF, which is fitted according to the equilibrium prediction. The resulting fit is seen in Fig. \ref{fig:potential_analysis_painel}c). Proper calibration of the transverse directions is achieved through independent measurements of the CCD's pixel size compared to a reference, while the longitudinal direction is measured by integrating the image brightness over the trapped sphere \cite{tebbenjohanns2019}, and calibrated via comparing the root-mean-square deviations of the matched simulation with the measured data. The reconstructed potential at the trapped sphere's centroid position -- obtained by taking the logarithm of the PDF -- is shown in Fig. \ref{fig:potential_recon}, with the corresponding parameters obtained from the experiment and in comparison to full Lorentz-Mie numerical simulation of a DFT with NA$=0.46$ in Table \ref{table}.

\begin{table}[]
\begin{tabular}{ccc}
\hline \hline
Parameter   & Experiment & Lorentz-Mie simulation  \\ \hline \
 $ k_{z} \SI[parse-numbers = false]{}{(\newton \per \meter)}$  & $(3.86 \pm 0.06)\times 10^{-7}$ & $(2.93 \pm 0.79)\times 10^{-7}$  \\
$k_{\rho z} \SI[parse-numbers = false]{}{(\newton \per \cubic \meter)}$  & $(8.81
 \pm 0.14)\times 10^7$ & $ (8.84 \pm  0.25)\times 10^7$ \\
$ k_{\rho} \SI[parse-numbers = false]{}{(\newton \per \cubic \meter)}$ & $(2.26 \pm 0.07)\times 10^8$ & $(1.63 \pm 0.17)\times 10^8$ \\ \hline \hline 
\end{tabular}
\caption{Reconstructed potential parameters in comparison to numerical simulations of Lorentz-Mie theory. Error bars are obtained by dividing the experimental and simulated data into five sets and taking the standard deviation.}
\label{table}
\end{table}



\textit{Conclusion.} -- In summary, we have experimentally investigated a structured light dark focus tweezer for dielectric microparticles immersed in a high refractive index medium. We have shown stable trapping and isolation from surrounding objects by repulsive optical forces, which induce a non-harmonic potential landscape. 

We expect the dark trap will find use both in applied and fundamental physics. In biophysics, dark tweezers can provide stable trapping for organisms with reduced laser heating. This is advantageous, as it has been shown that bright tweezers hinder cell
reproduction and exponentially decrease cell lifetime even at modest trapping powers \cite{Pilt2017}. 
Moreover, the dark focus tweezer
can be used for vacuum optical trapping using doped nanoparticles, for instance with rare-earth atoms \cite{shan2021optical} or Mie particles \cite{lepeshov2023levitated}.
%
Note that implementing a vacuum dark tweezer requires advancements in material science. Optical absorption by particles with internal resonances typically lead to unstable dynamics and particle loss caused by spectral and geometrical imperfections. 
If these challenges can be overcome, the dark focus tweezer could provide the advantage of a significantly reduced
internal bulk temperature of the particle, consequently reducing the decoherence effects caused by thermal emission.

\begin{acknowledgments}
We acknowledge Bruno Melo, Igor Brandão, Cyril Laplane and Lukas Novotny for helpful discussions, and Angela Duncke for help on the sample preparation protocol.
Conselho Nacional de Desenvolvimento Cient\'{\i}fico e Tecnol\'ogico (CNPq),
Coordena\c c\~{a}o de Aperfei\c coamento de Pessoal de N\'\i vel Superior (CAPES), 
Funda\c c\~{a}o Carlos Chagas Filho de Amparo \`{a} Pesquisa do Estado do Rio de Janeiro (FAPERJ), 
Instituto Nacional de Ci\^encia e Tecnologia de Informa\c c\~ao Qu\^antica (INCT-IQ 465469/2014-0),
Fundação de Amparo à Pesquisa do Estado de S\~ao Paulo (FAPESP, processos 2021/06823-5 and 2021/06736-5),
Code and data availability: GitHub. https://github.com/QuantumAdventures/optical-bottle-beam
\end{acknowledgments}

\bibliographystyle{apsrev4-1}
\bibliography{main.bib}


\onecolumngrid

\section*{Appendix A: The Dark Focus Tweezer} 

The Dark Focus Tweezer (also known as  optical bottle beam) used in this work consists in a superposition of a Gaussian and a Laguerre-Gauss (LG) mode with $\ell = 0, p \neq 0 $ with a relative phase of $\pi$. The LG mode functions read, 
 \begin{eqnarray}
  \label{eq:LGbeam2}
      \hspace*{-2em}&\,& u_{\ell,p}(\rho,\phi,z) =\sqrt{\frac{2}{\pi\omega(z)^2}}\sqrt{\frac{p!}{(\vert \ell \vert+p)!}}\times\nonumber\\&&\left( \frac{\sqrt{2}\rho}{\omega(z)} \right)^{\vert\ell\vert}\mathrm{L^{\vert\ell\vert}_p}\left( \frac{2\rho^2}{\omega(z)^2} \right)\exp\left[-\frac{\rho^2}{\omega(z)^2}\right]\times \exp[ik_mz+ik_m\frac{\rho^2}{2R(z)}-i\zeta(z)+i\ell\phi],
\end{eqnarray}
where $k_m$ is the wavenumber in the medium and $\omega(z)$, $R(z)$, $\zeta(z)$ and $\mathrm{L^{\vert\ell\vert}_p}$ 
are the beam width, the wavefront radius, the Gouy phase and the Associated Laguerre polynomial. These quantities are,
\begin{eqnarray}
\omega(z)&=&\omega_0\sqrt{1+\frac{z^2}{z^2_R}};\\
  R(z) &=& z\left(1+\frac{z_R^2}{z^2} \right);\\
\zeta(z)&=&(2p+\vert\ell\vert+1)\arctan \frac{z}{z_R};\\
    \mathrm{L_p^{\vert\ell\vert}}(x) &=& \sum^p_{i=0}\frac{1}{i!}\binom{p+\vert\ell\vert}{p-i}(-x)^i
\end{eqnarray}
\noindent where the Rayleigh range ($z_R$) and the beam waist ($\omega_0$) are defined as
\begin{eqnarray}
\omega_0=\frac{\lambda_0}{\pi \mathrm{NA}}\,,\quad z_R = \frac{n_{m}\lambda_0}{\pi \mathrm{NA}^2} \label{eq:waist_and_Rayleigh_range}
\end{eqnarray}
\noindent with $\lambda_0$ being the wavelength in vacuum, $n_{m}$ the medium refractive index and $\mathrm{NA}$ the numerical aperture. Throughout this work we consider linearly polarized fields.

\begin{figure}[h]
    \centering
    \includegraphics[ trim={5, 10, 5, 10},clip, width=0.5\textwidth]{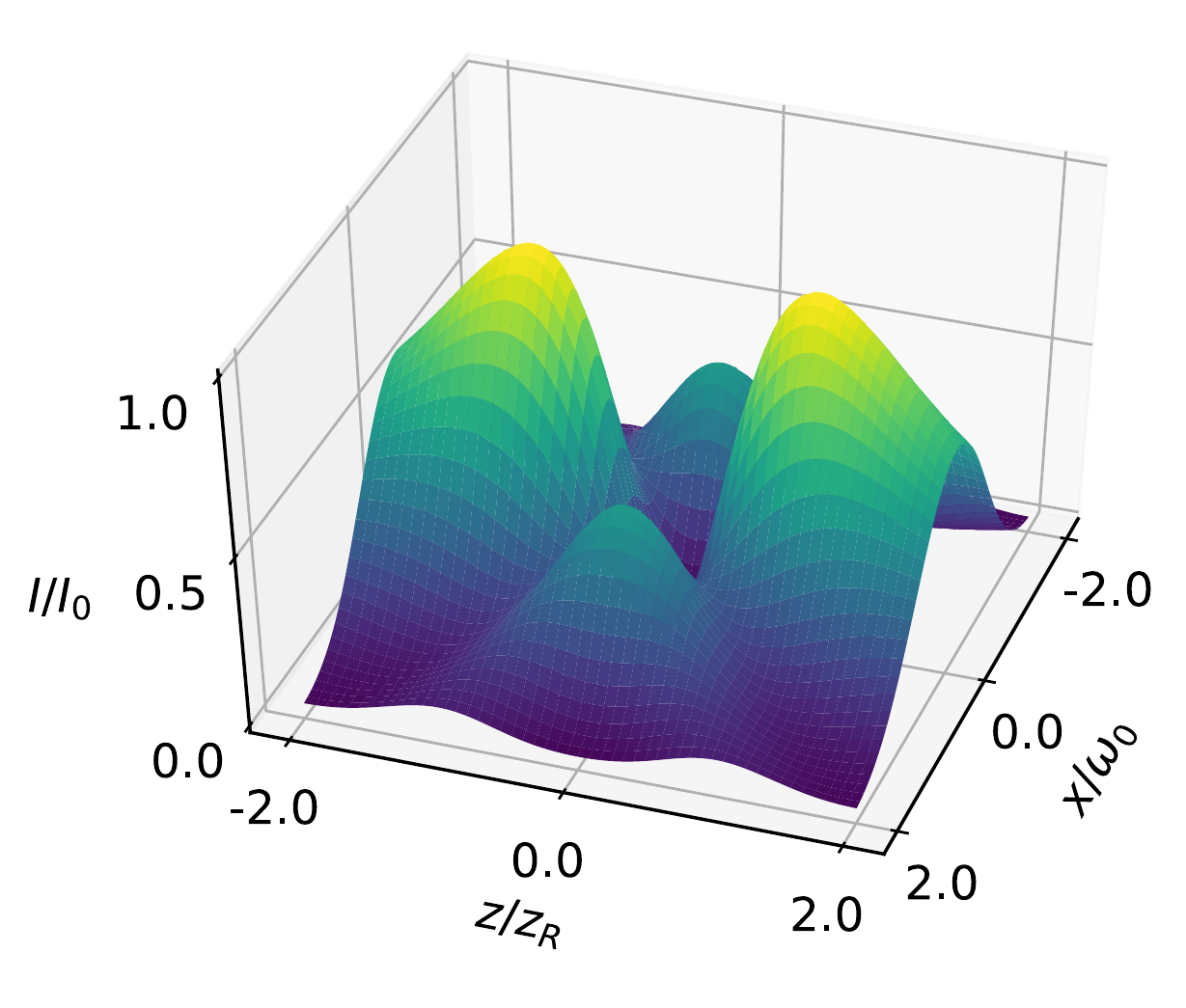}
    \caption{Intensity landscape of the dark focus tweezer generated by superposing a Gaussian and Laguerre-Gauss mode with $ \ell = 0, p = 1 $ with a relative phase of $ \pi$. The beam intensity is normalized by $I_0 = P_0/(\pi\omega_0^2)\,$.}
    \label{fig:OBB_intensity}
\end{figure}

The intensity profile of a Gaussian beam with total power $P_0$ is simply $I_G(\mathbf{r})=P_0\,|u_{0,0}(\mathbf{r})|^2$.
The resulting intensity profile of the DFT beam with the same total power $P_0$ reads,
\begin{eqnarray}
\label{eq:exact_intensity}
I_{\mathrm{B,p}}(\rho,z) &=& P_0\,\left|\frac{u_{0,0}(\mathbf{r}) - u_{0,1}(\mathbf{r})}{\sqrt{2}}\right|^2
\\
&=&\frac{P_0}{\pi\omega(z)^2}\exp\left[-\frac{2 \rho^2}{\omega(z)^2}\right]\times \bigg[1-\hspace{-0.5mm}2\cos\left(2p\arctan\frac{z}{z_R}\right) \mathrm{L^0_p}\left(\frac{2\rho^2}{\omega(z)^2}\right)+ \ \mathrm{L^0_p}\left(\frac{2\rho^2}{\omega(z)^2}\right)^2 \bigg]\,,
\nonumber
\end{eqnarray}
where $P_0$ is the total power of the beam. Fig. \ref{fig:OBB_intensity} shows the resulting normalized intensity profile of a DFT with $ p = 1$, which we adopt throughout this work.

Following \cite{melo2020a}, we define the width $W$ and height $H$ of the OBB as the distance between the two intensity maxima surrounding the dark focus along the $x$ axis ($z$ axis). These can be found by solving
\begin{eqnarray}
\label{eq:width}
dI_{\mathrm{B,p}}(x,0,0)/dx\vert_{x=W/2}=0\, , \ \ \ \ \ 
\label{eq:height}
dI_{\mathrm{B,p}}(z,0,0)/dz\vert_{z=H/2}=0\, ,
\end{eqnarray}
yielding $W=2\omega_0,H=2z_R$ for $p=1$. 
Note the width of the OBB scales as $\mathrm{NA}^{-1}$, while the height scales as $\mathrm{NA}^{-2}$.

The DFT intensity in the Rayleigh regime ($R \ll \lambda_{0} $) gives rise to a scattering (non-conservative) and gradient (conservative) force fields, respectively given by \cite{jones2015},
 \begin{eqnarray}
 \vec{F}_{{\mathrm{B,p}}}^{(s)}(\vec{r}) &=& \hat{z}\frac{128\pi^5R^6}{3c\lambda_0^4}\left(\frac{m^2-1}{m^2+2}\right)^2n_{m}^5I_p(\vec{r})\\
 \vec{F}_{{\mathrm{B,p}}}^{(g)}(\vec{r})&=&\frac{2\pi n_{m}R^3}{c}\left( \frac{m^2-1}{m^2+2}\right)\nabla I_p(\vec{r})
\end{eqnarray}

\noindent where $m=n_p/n_m$ is the particle-medium refractive index ratio and $I_{{\mathrm{B,p}}}(\vec{r})$ is the intensity distribution of the beam. 

The potential landscape associated to the gradient force near the origin is given by an approximate polynomial potential 
\begin{eqnarray}\label{eq:potential-polynomial}
    \frac{V_{{\mathrm{B,p}}}(\rho,z)}{V_0}\approx \mu' z^2 - \eta' \rho^2z^2 + \chi' \rho^4 
    \label{eq:app_potential}
\end{eqnarray}
with,
\begin{eqnarray}
    \mu' = \frac{4p^2}{z_R^2}  \ , \ 
 \eta' = \frac{8p^2(p+1)}{\omega_0^2z_R^2} \ , \  \chi' = \frac{4p^2}{\omega_0^4}
 \label{eq:coefficients_dipole}
\end{eqnarray}
and $V_0=[2\pi n_mR^3(m^2-1)/c(m^2+2)]I_0$. Numerical simulations of Lorentz-Mie scattering theory validate that in the intermediate regime, the same potential form \eqref{eq:app_potential} can be considered, despite the coefficients $ \mu', \eta', \chi' $ are no longer expressed as simple functions of the beam parameters.

\begin{figure}[h]
    \centering
    \includegraphics[width=1\textwidth]{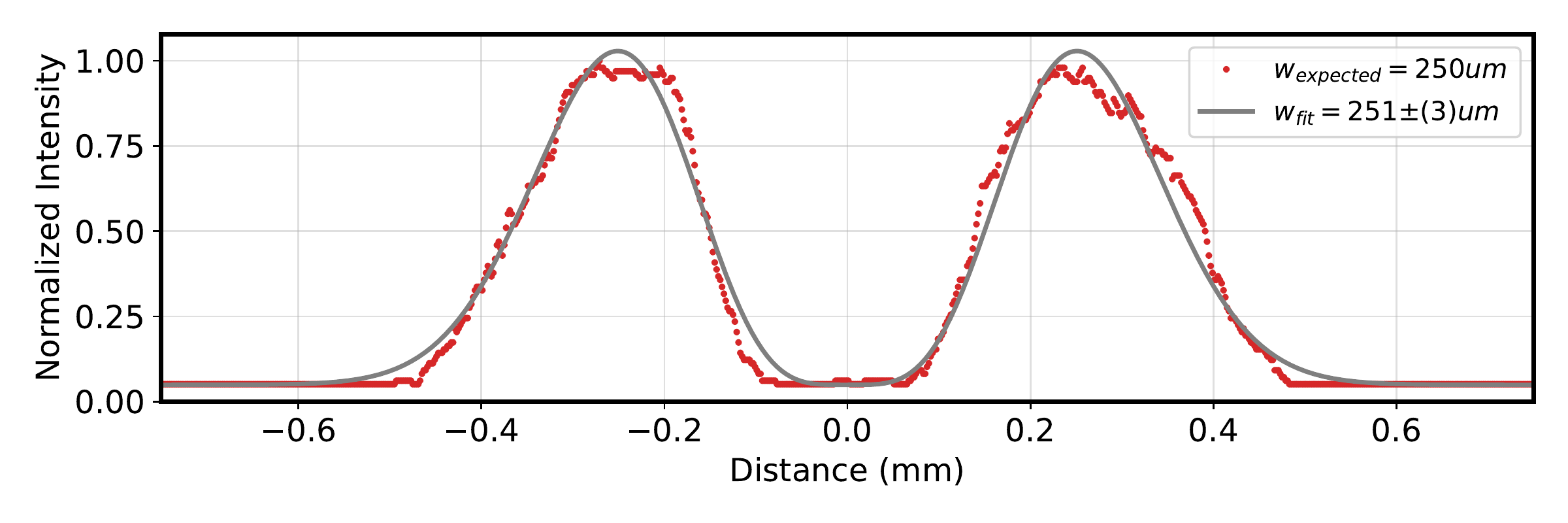}
    \caption{Intensity plot along the $ x $ direction of OBB transverse profile at the dark focus ($\theta = \pi$), with corresponding fit. Considering the optical setup (telescope after the SLM), a profile beam waist of $w_{\mathrm {expected}}= \SI{250}{\mu m } $ is expected. The beam radius obtained by fitting \eqref{eq:exact_intensity} to the transverse profile along the $ x $ direction is $ w_{\rm fit}=\SI[separate-uncertainty = false]{251 \pm 3}{\mu m}$,  demonstrating the generation of a good quality OBB.}
    \label{fig:focal_plane_fit}
\end{figure}

As explained in the main text, we generate the OBB using an SLM. Figure \ref{fig:focal_plane_fit} depicts the beam profile at the focal plane, together with a corresponding fit of the beam intensity according to \eqref{eq:exact_intensity}. We observe excellent agreement between the expected beam profile and the mode generated by the SLM.

\section*{Appendix B: Effective Potential Validation} 

As discussed in the main text, the potential in the dipole regime can be expanded as a polynomial function of $\rho$ and $z$. From Eq. \eqref{eq:potential-polynomial}, it is possible to find expressions for the forces in the $x$, $y$ and $z$ direction. To validate the quartic polynomial model in the intermediate regime, we simulated the optical forces considering particles of different radii using the MatLab toolbox described in \cite{nieminen2007}. We fit a polynomial to the force vectors $f_x$, $f_y$ and $f_z$ obtained from the simulation, returning estimations for each of the components, namely $\hat{f}_x$, $\hat{f}_y$ and $\hat{f}_z$. The quality of the fit in each axis can be evaluated using a root-mean-squared error (RMSE) divided by the root-mean-square force.
Finally an average RMSE over each axis is considered,
\begin{equation}\label{eq:rmse_avg}
    \mathrm{RMSE_{avg}} = \frac{1}{3}\sum_{i\in\{x,y,z\}}\Bigg\{\sqrt{\frac{\sum\limits_{j=1}^{N}(f_i^j-\hat{f}_i^j)^2}{\sum\limits_{j=1}^N f_i^{j\,2}}}\Bigg\},
\end{equation}
where $N$ represents the number of points used for discretization during simulation.

\begin{figure}[h]
    \centering
    \includegraphics[width=0.5\textwidth]{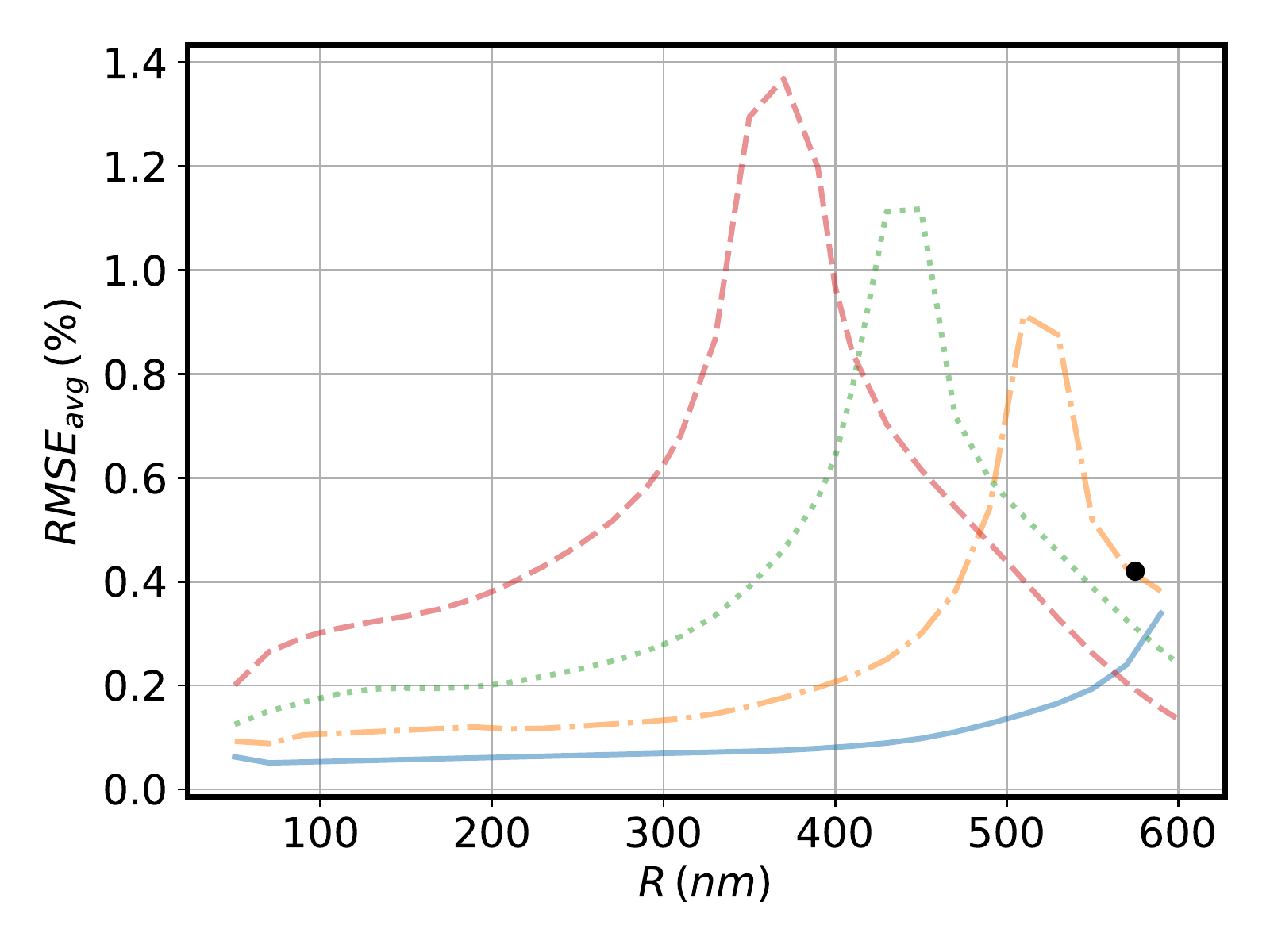}
    \caption{$\mathrm{RMSE_{avg}}$ of the polynomial fit for the optical forces as a function of the particle radius. Simulations were executed considering NA = 0.40 (\protect\tikz[baseline]{\protect\draw[line width=0.25mm] (0,.8ex)--++(0.5,0) ;}),
	 NA = 0.46  (\protect\tikz[baseline]{\protect\draw[line width=0.25mm,dash dot] (0,.8ex)--++(0.5,0);}), 
	 NA = 0.52 (\protect\tikz[baseline]{\protect\draw[line width=0.25mm,dotted] (0,.8ex)--++(0.5,0) ;}) and NA = 0.58
	 (\protect\tikz[baseline]{\protect\draw[line width=0.25mm,densely dashed] (0,.8ex)--++(0.45,0);}). The conditions in which the experiment was conducted are represented by the black circle ($\bullet$). 
    }
    \label{error_analysis}
\end{figure}

Fig. \ref{error_analysis} shows the $\mathrm{RMSE_{avg}}$ as a function of particle radius $ R $. Different values of NA were considered to ensure the approximation is valid under variations of the trapping beam focusing. The maximum error encountered is always less than $1.50\%$, for a particle radius of \SI{350}{nm} and an NA of $0.58$. For the experimental conditions described in the main text, the error is $0.416\%$, validating the quartic potential model within the experimental parameters. 

\section*{Appendix C: Probe beam calibration} 

\begin{figure}[ht!]
    \centering
    \includegraphics[width=0.5\textwidth]{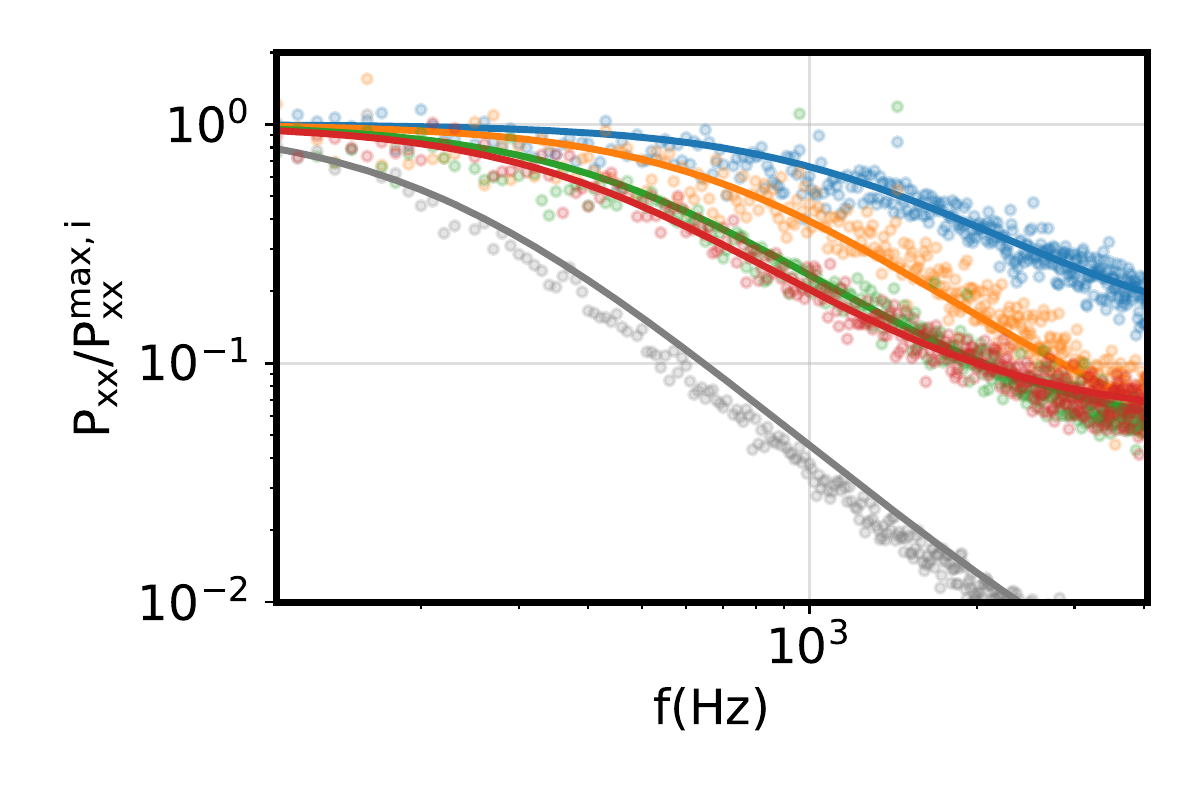}
    \caption{Probe beam calibration: PSD of probe beam scattering at different intensities for a particle trapped at fixed power $P_{T}=$ \SI{39}{mW}, with corner frequency  $f_c=$ \SI[parse-numbers = false]{(428.8 \pm 14)}{Hz} (red).  The probe beam power $ P_{P}^{i} $ ($ i = 1,2,3 $) is progressively decreased: $P_P^{1}=$ \SI{91}{mW}, $f_c=$ \SI[parse-numbers = false]{(1272.2 \pm 40) }{Hz} (blue);  $P_P^{2}=$ \SI{45}{mW}, $f_c=$\SI[parse-numbers = false]{(770.8 \pm 19)}{Hz} (orange); $P_P^{3}$ = \SI{19}{mW}, $f_c=$ \SI[parse-numbers = false]{(492.0 \pm 22)}{Hz} (green). For comparison, the PSD of the probe beam in the absence of a trap is also shown (grey). 
    Probe beam powers below \SI{19}{mW} allows for position read-out without significant disturbance to the trap. Each PSD curve is normalized by its corresponding plateau value $\mathrm{P_{xx}^{max,i}}$ obtained from the Lorentzian fit. 
    }
    \label{fig:psd_section}
\end{figure}

To determine the proper alignment and optimal power of the probe beam we perform standard Gaussian tweezer experiments at various probe beam powers with a particle immersed in aqueous solution at a fixed trap power of $ P_{T} = \SI{39}{mW}$. As the probe power is decreased, the measured PSD approaches that of the trapping beam alone. 
The power at which the probe and trap PSDs are indistinguishable determines the power at which the probe beam causes negligible influence on the trap within experimental uncertainties; this power was measured to be $ P_{P} \leq \SI{19}{mW} $. 
The result of this calibration measurement is shown in Fig. \ref{fig:psd_section}a). See the caption for details.

\section*{Appendix D: Absorbed power in the dark focus and trap performance} 

To quantify the energy absorption in the dark focus in comparison to a standard bright tweezer we turn to the formula for the absorbed power in terms of the beam's and particle's parameters \cite{ricci2022chemical},
\begin{eqnarray}
P_{\mathrm{abs}} = 12\pi I_0 \frac{V}{\lambda} \mathfrak{I}\left(\frac{\epsilon - 1}{\epsilon + 2}\right)
\end{eqnarray}
where $I_0$ is the light intensity and $A_p \equiv V/\lambda$ is the effective cross-section of the particle, so that $I_0 V/\lambda$ represents the optical power effectively seen by the particle. A meaningful comparison between the usual Gaussian bright tweezer and the proposed dark focus tweezer can be built from the following \textit{absorption ratio},
\begin{eqnarray}
\eta_{\mathrm{abs}} = \frac{P^{\,\prime}_{\mathrm{abs}}}{P_{\mathrm{abs}}} =
\frac{P_{B}}{P_{G}} \times 
 \frac{\int_{A_p} \vert u_B(\mathbf{r})\vert^2  d^2\mathbf{r}}{\int_{A_p} \vert u_G(\mathbf{r}) \vert^2 d^2\mathbf{r}}
\end{eqnarray}
where $u_G$ is the Gaussian field distribution used in a standard bright tweezer, $u_B$ is the optical bottle beam (OBB) field distribution used in the dark focus tweezer and $ P_{G}, P_{B} $ are the powers in the Gaussian and dark focus traps, respectively. Note that the only free parameters in the definitions of the modes $ u_{B} $ and $ u_{G} $ are the total power and the beam waist $w_0$, and both $u_B$ and $u_G$ are normalized with respect to integration over the whole transverse plane, while their intensity distributions are completely different. 

In order to give a quantitative estimation of the ratio $\eta_{\mathrm{abs}}\,$, we assume that the absolute value of the polarizability of the particle immersed in the surrounding medium is the same in both the dark focus and Gaussian traps. The radius of the Silica particles used in our experiment is $R=\SI{575}{nm}$ and the laser wavelength is $\lambda = \SI{780}{nm}\,$, giving $A_p = 1.02 \times 10^{-12} \ \mathrm{m}^{2}$.

\begin{figure}[h]
    \centering
    \includegraphics[width=0.5\textwidth]{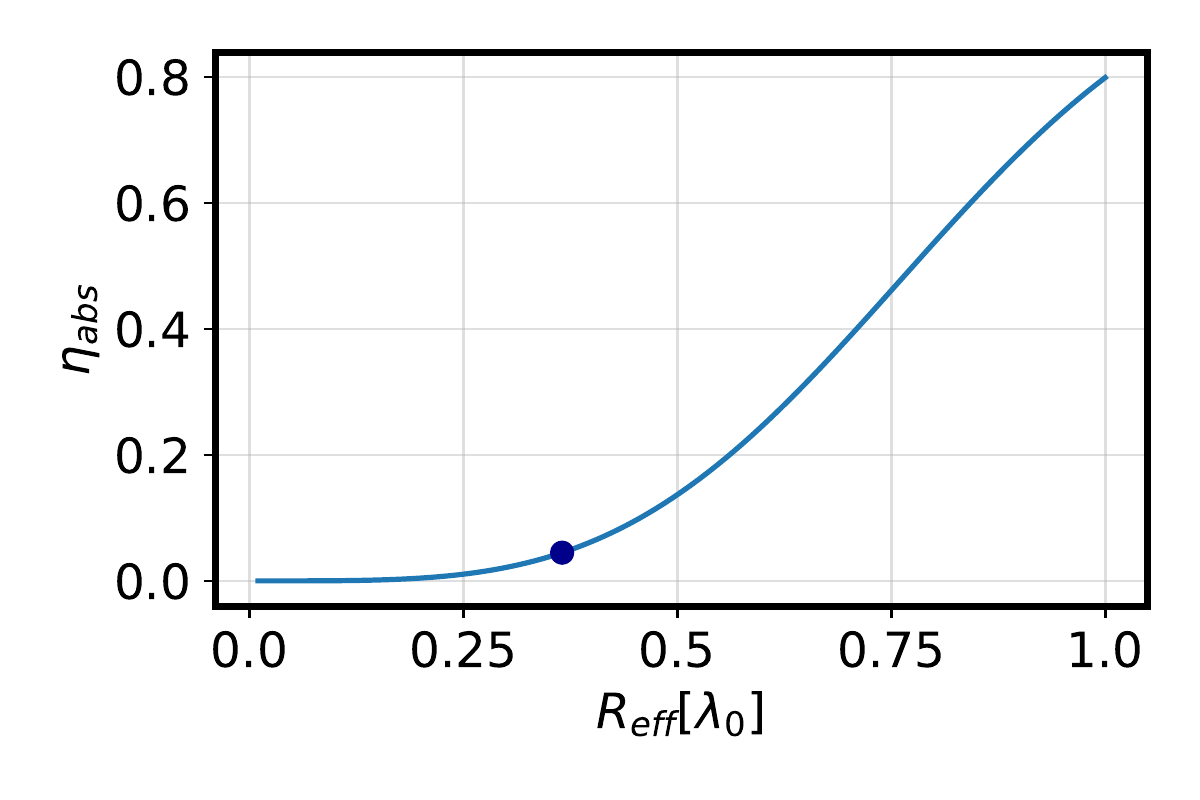}
    \caption{Ratio $\eta_{\mathrm{abs}}$ as a function of the effective particle radius $ R_{\mathrm{eff}} $.  }
    \label{fig:eta_x_radius}
\end{figure}

If we set the waists and total power of the dark focus and Gaussian tweezers to be the same, that is $ P_{B} = P_{G} $,
using the above parameters for particle radius, wavelength $\lambda$, cross-section $ A_{P} $ and the mode functions employed in the experiment, we find an absorption ratio of $\eta_{\mathrm{abs}}=0.045$. 
Moreover, the value of $ \eta_{\mathrm{abs}} $ decreases with the particle radius, as expected for the dark focus tweezer. The numerical value of $\eta_{\mathrm{abs}}$ as a function of effective particle radius $ R_{\mathrm{eff}} = \sqrt{A_{p}/4\pi}$ is indicated in Figure \ref{fig:eta_x_radius}.

\noindent It is interesting to compare the dark focus trapping characteristics to that of a standard Gaussian tweezer. Note that the trapping potential associated with $u_{B}$ has a \textit{quartic} dependence with the radial coordinate ($V \sim \frac{k_{\rho}}{4}\rho^{4}$) over the transverse plane, so it is not suitable to compare $k_{\rho}$ with the spring constant of the quadratic potential produced by $ u_{G}$.
Once the powers $P_G$, $ P_{B} $ and beam waist $w_0$ are fixed for both beams, however, a comparison between the trapping potentials can be made through the corresponding trap depths.

\begin{figure}[h!]
    \centering
    \includegraphics[width=0.9\textwidth]{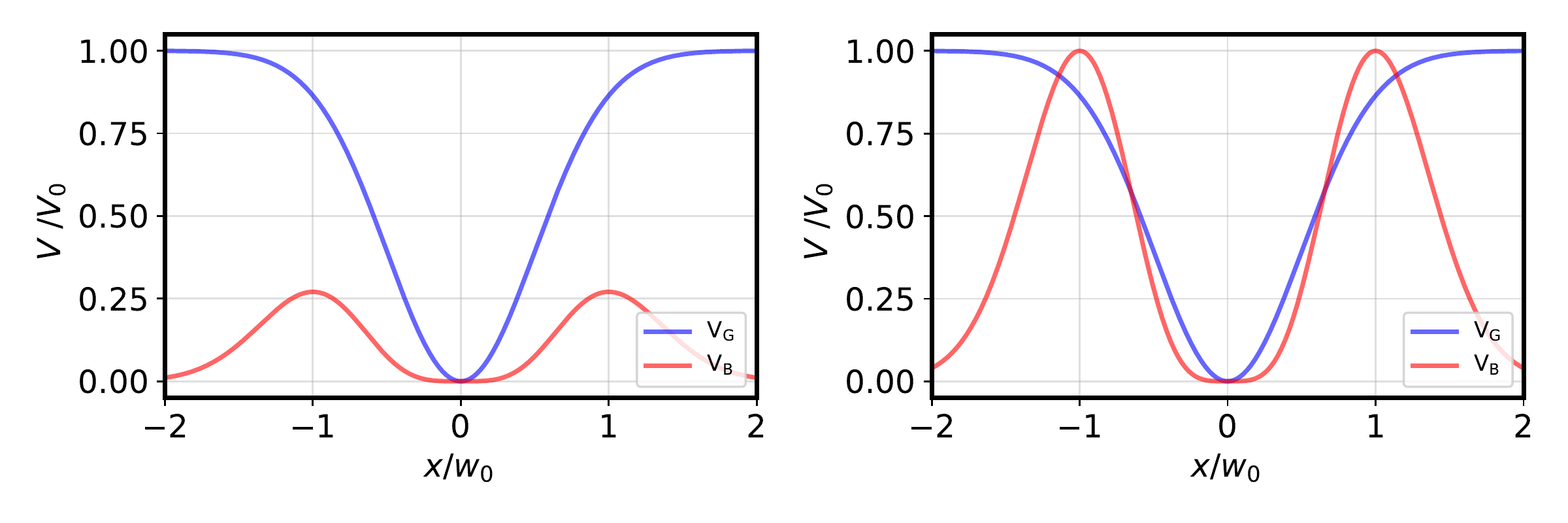}
    \caption{Comparison between the dark focus and Gaussian tweezers for the transverse directions and corresponding absorption ratios. \textbf{Left:} 
    For the same optical power and waist, the dark focus presents an absorption ratio of $ \eta_{\mathrm{abs}} = 0.045 $, i.e. the particle absorbs about $ 4.5\%$ of the power absorbed in a standard optical trap. This reduction in absorption comes at the expense of a reduction in trap depth of $V_{0,B} \approx 0.27 V_{0,G}$. \textbf{Right:} Increasing the dark focus power to about a factor of 4 the trap depth of both tweezers can be matched, yielding equivalent trap performances with an absorption ratio of $ \eta_{\mathrm{abs}} \approx 0.17 $. This is a substantial reduction in absorption, i.e. $ \approx 17\% $ of the total absorption in a standard Gaussian trap. In both plots, the zero potential energy reference was set so that the two minima coincide. }
    \label{fig:V_G_and_V_OBB_x}
\end{figure}

The reduction in absorption in the dark focus trap compared to the absorption in a Gaussian tweezer of the same power and waist is accompanied by a reduction in trap depth along the transverse plane, as can be seen in the Left plot in Figure \ref{fig:V_G_and_V_OBB_x}. 
A straightforward calculation using the mode functions shows that, 
\begin{eqnarray}
V_{0,B} = \frac{2}{e^{2}}V_{0,G} \approx 0.27 V_{0,G}
\end{eqnarray} 
where $ V_{0,B}, V_{0,G} $ are the trap depths in the dark focus and Gaussian tweezers, respectively. Therefore, we see that a significant 20-fold reduction in absorbed power ($\eta_{\mathrm{abs}} = 0.045$) comes at the expense of a 4-fold reduction in trap depth. Note that in our experiment the observed trap depth was on the order of $ V_{0,B} \approx 100 k_{B} T $, guaranteeing that the particle was well within the trapping potential. 

Alternatively, we can set the modes $ u_{B} $ and $ u_{G} $ at different powers such that their trap depths are equal, as shown in the Right plot in Figure \ref{fig:V_G_and_V_OBB_x}. In this case, the power in the dark focus has to be approximately 4 times larger than in the Gaussian beam, resulting in an approximately 4-fold increase in absorption. This yields an absorption ratio of $ \eta_{\mathrm{abs}} \approx 0.17 $, showing that the trap performance of a standard Gaussian beam can be matched by the dark focus tweezer while at the same time achieving a substantial reduction in absorbed power, i.e. $ \approx 17\% $ of the total absorption in a standard Gaussian trap.

\begin{figure}[h]
    \centering
    \includegraphics[width=0.9\textwidth]{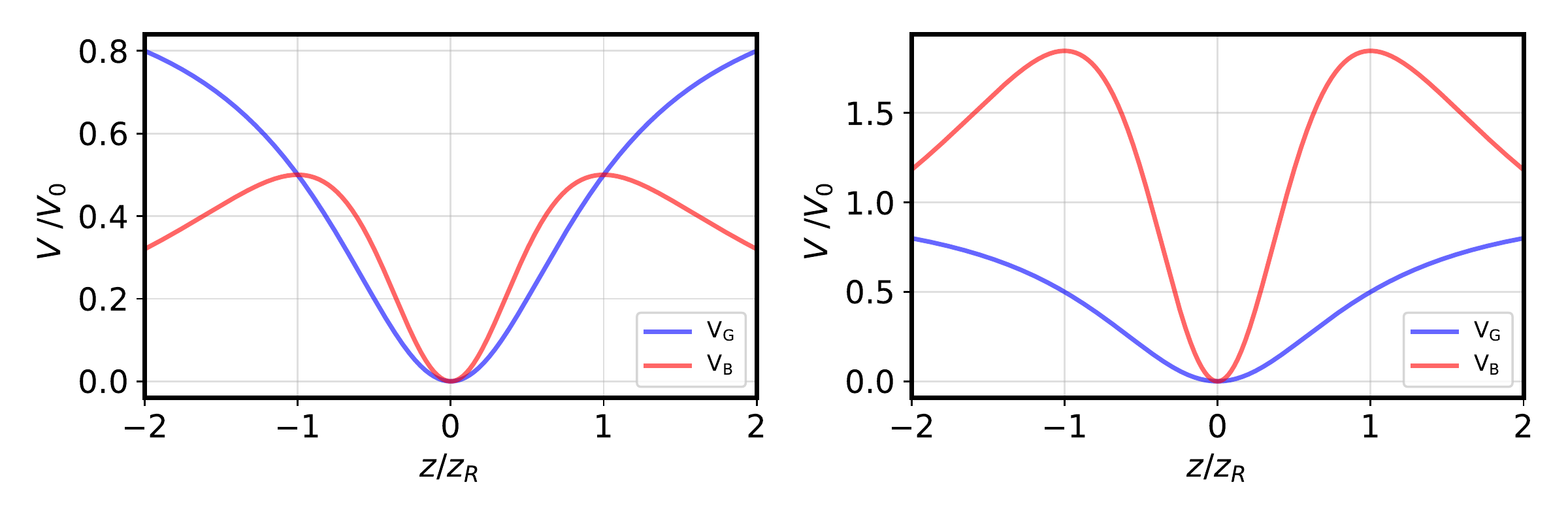}
    \caption{Comparison between the longitudinal trapping potentials in the dark focus ($ V_{\mathrm{B}}$) and Gaussian tweezer ($V_{\mathrm{G}}$) with equal waists. \textbf{Left:} The two types of trapping beams are assumed to have equal power. The dark focus presents a reduction of about $50\%$ in trap depth, however, near the origin it provides a quadratic potential twice stiffer than a Gaussian trap. \textbf{Right:} The dark focus beam power is assumed to be about 4 times larger than the Gaussian beam, in order to match the transverse potential depths. Besides being stiffer, the dark focus longitudinal trapping potential is approximately twice deeper than the Gaussian one.}
    \label{fig:V_G_and_V_OBB_z}
\end{figure}

Concerning the longitudinal direction, the trapping potentials of both the bottle and Gaussian beams have a quadratic dependence with the $ z $ coordinate to leading order, so the spring constants can be directly compared. As shown in the Left plot in Figure \ref{fig:V_G_and_V_OBB_z}, a dark focus with the same power as the Gaussian tweezer presents a reduction of about $ 50\% $ in trap depth, however, presenting a stiffer potential near the origin. From Eq. (13) of \cite{melo2020b}, the dark focus trap longitudinal spring constant is twice larger than that of a Gaussian beam of equal power. Increasing the power of the dark focus by a factor of four to match the transverse trapping characteristics of a Gaussian beam, as discussed above and shown in the Right plot in Figure \ref{fig:V_G_and_V_OBB_z}, the longitudinal potential trap depth in the dark focus becomes approximately twice that of the Gaussian.





\end{document}